\documentclass[journal]{IEEEtran}
\usepackage{cite}
\usepackage{amsmath,amssymb,amsfonts,amsthm}
\usepackage{float, array}
\usepackage{algorithmic}
\usepackage{graphicx}
\usepackage{textcomp}
\usepackage{xcolor}
\usepackage{wrapfig}
\usepackage{longtable}
\usepackage{pdflscape}
\usepackage{subcaption}
\usepackage{url}
\def\diag{{\rm diag}}
\def\vec{{\rm vec}}

\def\BibTeX{{\rm B\kern-.05em{\sc i\kern-.025em b}\kern-.08em
T\kern-.1667em\lower.7ex\hbox{E}\kern-.125emX}}
\begin{document}
\title{Exploring the Synergy:\\ A Review of Dual-Functional Radar Communication Systems}

\author{Ali Hanif, {\em Graduate Student Member, IEEE},
        Sajid Ahmed, {\em Senior Member, IEEE}, Mohamed-Slim\\ Alouini, {\em Fellow, IEEE},
         and Tareq Y. Al-Naffouri, {\em Senior Member, IEEE}}

\maketitle
\begin{abstract}
This review paper examines the concept and advancements in the evolving landscape of Dual-functional Radar Communication (DFRC) systems. Traditionally, radar and communication systems have functioned independently, but current research is actively investigating the integration of these functionalities into a unified platform.  This paper discusses the motivations behind the development of DFRC systems, the challenges involved, and the potential benefits they offer. A discussion on the performance bounds for DFRC systems is also presented. The paper encompasses a comprehensive analysis of various techniques, architectures, and technologies used in the design and optimization of DFRC systems, along with their performance and trade-offs. Additionally, we explore potential application scenarios for these joint communication and sensing systems, offering a comprehensive perspective on the multifaceted landscape of DFRC technology.
\end{abstract}

\begin{IEEEkeywords}
Joint Communication and radar/radio Sensing (JCAS), Dual-functional Radar Communications (DFRC), Integrated Sensing and Communications (ISAC), Wireless Communications, and Radar.
\end{IEEEkeywords}

\section{INTRODUCTION}
\subsection{Background and Motivation}
\IEEEPARstart{T}{he} number of worldwide Internet of Things (IoT) and non-IoT connections have been forecasted to reach over 30 billion and 10 billion respectively by 2025 \cite{1}. Net proceeds of the C-band 5G auction in the US amount to \$81.1 billion in 2021 \cite{2}. With such a drastic increase in the number of users and the cost of the spectrum, there is an increased research interest in the optimal utilization of the frequency spectrum. Wireless communications and radar sensing are two fields that have traditionally been progressing independently for decades. The first operational use of radar dates back to WW-II when the ‘Chain Home’ radar network was employed by the Royal Air Force to detect enemy aircraft \cite{44}. Since then, certain portions of the frequency spectrum have been used exclusively by radars for various operations like airborne radars, Air Traffic Control (ATC) radars, weather radars, ground surveillance radars, marine radars,...etc. Thus, spectrum sharing between communications and radars can help in solving the spectrum congestion problem leading to a better spectral efficiency. Moreover, joint design of radar and communication systems can offer other important benefits such as reduced power consumption,  form factor reduction, cost saving, beamforming efficiency, and performance improvement of both functions via cooperation \cite{hassanien2019dual,book2022joint}.

Earlier research to solve the spectrum congestion problem focused on the coexistence of radar and communication systems where both systems can operate simultaneously in a cooperative fashion without interfering with each other. Radar-Communication Coexistence (RCC) approaches include \cite{liu2020joint} opportunistic spectrum access \cite{wang2008application}, interference channel estimation \cite{mahal2017spectral,joint}, precoder design \cite{babaei2013nullspace} and optimization based design \cite{liu2017robust,codesign,2019joint}. However, the efficacy of these interference cancellation techniques dictates strict requirements on the mobility of nodes and information exchange, thus, limiting the improvement in spectral efficiency  \cite{book2022joint}. Another major drawback of coexistence approaches is that for efficient cooperation, radar and communication systems have to exchange side information such as the channel state information, modulation formats,...etc. \cite{liu2018toward}. This exchange is usually made possible by a control center connected to both systems \cite{joint}, thus, considerably increasing the overall complexity of the system.

Research on joint Communications and Radar (C\&R) systems has been reported under various names such as Radar-Communications (RadCom), Joint Communications (Radar)-Radar (Communications) (JCR/JRC), Joint Communication and radar/radio Sensing (JCAS), Dual-functional Radar Communications (DFRC), and Integrated Sensing and Communications (ISAC). The interference issue of coexistence systems can be tackled by using the same signal for communications and radar sensing. DFRC system refers to the joint systems that use a single transmit waveform for both functions \cite{liu2020joint,hassanien2019dual,congest2019dual}. The first ever published DFRC scheme was proposed in \cite{mealey1963method}, in which the communication bits were coded on the radar pulse interval\cite{liu2020joint}. A detailed timeline of the evolution of DFRC techniques can be found in \cite{liu2020joint}.
\subsection{Classification of JCAS Systems}
JCAS systems can be classified into three broad categories \cite{book2022joint}.
    \subsubsection{Communication-centric Design or JCR} In JCR systems, design priority is on communications and radar functionality is added as a secondary feature \cite{kumari2017ieee,rahman2019framework,kumari2019adaptive}. Communication waveforms are exploited to extract sensing information from target echoes. Communication performance usually remains unaffected, however, sensing performance needs optimization and is scenario-dependent. Accurate and efficient sensing parameters estimation algorithms, resolution of clock asynchrony, and full duplex operation are the key research problems here for improving the performance of radar functionality \cite{hassanien2019dual}. An important advantage of these systems is in location-aware services and applications like V2X (Vehicle-to-Everything) network, where they can provide sensing-assisted communications and considerably reduce the beam training or channel state information estimation overhead \cite{gonzalez2016radar,liu2020radar}.
   \subsubsection {Radar-centric Design or JRC} In JRC systems, design priority is on radar sensing and communication information is embedded in known radar waveforms over time, frequency, code, or spatial domains \cite{congest2019dual,rossler2011software,wu2020waveform}. Index modulation (IM) can also be employed where information is represented by indexes of antennas, frequency or code and there is no waveform modification involved \cite{huang2019dual}. Radar performance usually remains unaffected, however, communication performance is limited by achievable data rates. One advantage is that long-range communication can be achieved owing to the high transmission power of radar systems.
    \subsubsection {Dual-functional Radar Communications (DFRC)} In DFRC systems, the two functions are jointly designed and optimized from the start and they are not biased towards communications or radar functionality. They offer a tunable trade-off between the two functions and aim at meeting the requirements of the desired application. There is more freedom of design in this category as it's not limited by existing radar or communications systems.

In the rest of the paper, we will use the term JCAS to refer to generic joint C\&R systems and the term DFRC to refer to the joint design and optimization of dual-functional systems capable of performing both functions simultaneously utilizing common waveform and hardware.

\subsection{Comparison of Communications and Radar Sensing}
To better understand the joint design of C\&R systems, first, we need to compare them to have a look at their similarities and differences. Following is a comparison between these two functions in terms of signal waveforms, transmission power, bandwidth, and other specifications \cite{book2022joint}.
\subsubsection{Signal Waveforms} Radars employ waveforms that are usually simple, unmodulated single-carrier pulsed or Continuous-Wave (CW) signals occupying large bandwidth. It is desirable to have radar waveforms that have an ambiguity function with steep and narrow mainlobes. In contrast, communication signals usually have a complicated signal structure with advanced modulations comprising unmodulated (training/pilot) and modulated data symbols. These signals can be discontinuous over domains of space, time, and frequency.
\subsubsection{Transmitter Power} The transmission power is usually high for pulsed radars to support long-range operation and comparatively lower for short-range CW radars. Communication systems normally operate on less power since the coverage range is quite less as compared to radars. Radar signals aim to have a low peak-to-average-power ratio for efficient power amplification to achieve long-range operation.
\subsubsection{Bandwidth} In pulsed radars, short pulses of large bandwidths are transmitted, followed by a silent period for the reception of the targets' echoes. CW radars transmit continuous waveforms, such as chirp signals, typically covering a large range of frequencies. Large bandwidth is desired in radar signals to improve the range resolution. On the other hand, communication systems usually operate in a bandwidth-restricted scenario and communication signals have much smaller bandwidths as compared to radar waveforms.
\subsubsection{Frequency of Operation} Operational frequency bands for radars include L-band (1-2 GHz), S-band (2-4 GHz), C-band  (4-8 GHz), X-band (8-12 GHz), K$_u$-band (12-18 GHz), K-band (18-27 GHz), and K$_a$-band (27-40 GHz). Automotive radars operate in the W-band (usually 75-81 GHz). Communication frequencies lie in sub-6 GHz and mmWave bands. Terahertz frequencies are being employed for 6G.
\subsubsection{Duplex Operation} Pulse radars operate in half-duplex mode while CW radars operate in full-duplex mode. Communication systems operate either in Time Division Duplex (TDD) or Frequency Division duplex (FDD) modes.
\subsubsection{Clock Synchronization} In radars, transmitter and receiver clocks are synchronized to avoid ambiguity in sensing parameters' estimation. In the case of communications, colocated nodes may share the same clock but non-colocated nodes usually don't share the same clock.
\subsubsection{Performance Metrics} Different performance metrics are used to evaluate the performance of radars and communication systems. For the case of radars, important metrics are probabilities of detection and false alarm, maximum range, maximum velocity, peak-to-sidelobe level ratio, Signal-to-Clutter Ratio (SCR), and resolution (range, Doppler and angular). Important metrics for wireless communications are  Signal-to-Interference-plus-Noise Ratio (SINR), Bit Error Rate (BER), outage probability, capacity, latency and throughput.

The majority of the research about joint communication and sensing is focused on the first two categories of JCAS systems and there is less work addressing the last category of DFRC systems. In this paper, we will present a detailed review of the DFRC systems. The paper is organized as follows: The system model for a generic DFRC system is presented in Section II. Section III presents different performance bounds for DFRC systems. Section IV provides a detailed review of the different design aspects of DFRC systems while Section V discusses key potential application scenarios of the JCAS systems. Resource allocation in DFRC systems is discussed in Section VI and Section VII highlights associated challenges and future research directions. The paper is concluded in Section VIII.

\textbf{Notations:} Bold lower case letters (${\bf x}$) denote vectors and upper case letters (${\bf X}$) represent matrices. The $(p,q)$th element of a matrix ${\bf X}$ is represented by $x_{p,q}$. The superscripts $(\cdot)^s$ and $(\cdot)^c$ will be used to represent sensing and communication signals, respectively.  ${\bf 1}_N$ represents an $N\times 1$ vector of ones and the identity matrix of dimension $N\times N$ is denoted by ${\bf I}_N$. Transpose and conjugate transposition of a matrix are denoted by $(\cdot)^T$ and $(\cdot)^H$, respectively. The conjugate of a scalar is denoted by $(\cdot)^*$. $\vec{(\cdot)}$ represents the vectorization operation and $\mathbb{E}{(\cdot)}$ is the expectation operator.

\section{System Model}
In this section, we present a system model for a DFRC platform, aiming to provide the reader with a better understanding of the subsequent discussions presented in the following sections. The model presented herein is of a generic nature, and for details of specific system models pertaining to individual works, the reader is directed to the relevant references. In the context of DFRC, we consider a platform equipped with a Uniform Linear Array (ULA) of $N_T$ transmit and $N_R$ receive antennas. The inter-element spacing between adjacent antennas is half of the carrier's wavelength. The system serves $K$ communication users by transmitting $L$ symbols per user and sensing $Q$ targets simultaneously. The transmitted waveforms from different antennas can be represented as a vector ${\bf x}(t)= \begin{bmatrix}x_1(t) & x_2(t) & \ldots & x_{N_T}(t) \end{bmatrix}^T$ for $0\leq t\leq T_p$, where $T_p$ is the pulse-repetition interval or symbol period.

\subsection{Received Signal Model for Sensing}
For the purpose of sensing, the transmit and receive antennas are co-located. The received signal by the target located at a distance $R$ in the direction $\theta$, ignoring path loss, can be expressed as
\begin{eqnarray}
r^s(t) =  {\bf a}_T^T(\theta){\bf x}(t-\tau),  \label{eq:RxSignalModel}
\end{eqnarray}
where ${\bf a}_T(\theta)=\begin{bmatrix} 1 & e^{j\pi \sin(\theta)} & \cdots & e^{j(N_T-1)\pi \sin(\theta)}\end{bmatrix}^T$ is the transmit steering vector and $\tau = \frac{R}{c}$. The received power at the target can be written as \cite{SajidBernie_TSP11,SajidSlim_TSP14,SajidNaffouri_TSP17}
\begin{eqnarray}\label{eq:bp}
P(\theta, {\bf R}) &=& \mathbb{E}\{{\bf a}_T^T(\theta){\bf x}(t-\tau){\bf x}^H(t-\tau){\bf a}_T^*(\theta)\}, \notag\\
                   &=& {\bf a}_T^H(\theta){\bf R}{\bf a}_T(\theta), \notag
\end{eqnarray}
where ${\bf R} = \mathbb{E}\{{\bf x}(t-\tau){\bf x}^H(t-\tau)\}$ is the $N_T\times N_T$ sample covariance matrix of the waveforms. The signal model in \eqref{eq:RxSignalModel} can be used for beamforming and transmitting information symbols \cite{10205519}. The transmitted signal reflects back from the target and the corresponding signals at the $N_R$ receive antennas in vector form can be written as
\begin{eqnarray}
{\bf z}^{s}(t) = \beta e^{j2\pi f_{d} t} {\bf a}_R(\theta){\bf a}_T^T(\theta){\bf x}(t-2\tau) + {\bf v}^{s}(t). \label{RxSig1}
\end{eqnarray}
Here, $\beta$ is the reflection coefficient of the target, $f_{d}$ is the Doppler frequency induced due to the motion of the target of interest, and ${\bf z}^{s}(t) = \begin{bmatrix} z^s_1(t) & z_2^s(t) & \cdots & z_{N_R}^s(t)\end{bmatrix}^T$. ${\bf a}_R(\theta) = \begin{bmatrix} 1 & e^{j\pi \sin(\theta)} & \cdots & e^{j(N_R-1)\pi \sin(\theta)}\end{bmatrix}^T$ represent the receive steering vector, and ${\bf v}^{s}(t) = \begin{bmatrix} v_1^s(t) & v_2^s(t) & \cdots & v_{N_R}^s(t)\end{bmatrix}^T$ contains the contribution of noise, interfering signals from other targets, and clutter. The noise  at the $n$th receive antenna can be written as
\begin{equation}
v_n^s(t) = \sum_{q=1}^{Q-1} \beta_q e^{j2\pi f_d^q t} e^{j\pi (n-1)\sin(\theta_q)}{\bf a}_T^T(\theta_q){\bf x}(t-2\tau_q) + w_n(t) \notag.
\end{equation}
The received signal in \eqref{RxSig1}
is sampled from $N_R$ antennas and can be written in matrix form as
\begin{eqnarray}
{\bf Y}^s = \beta {\bf a}_R(\theta){\bf a}_T^H(\theta){\bf X}_{l_\tau} {\cal D}(f_d) + {\bf V}^s. \label{eq:RxSig}
\end{eqnarray}
Here, ${\bf Y}^s = \begin{bmatrix}{\bf y}_1^s & {\bf y}_2^s & \cdots & {\bf y}_{N_R}^s\end{bmatrix}^T \in {\cal C}^{N_R\times L}$ while ${\bf y}_n^s =\begin{bmatrix} {y}_n^s(0) &  y_n^s(1) & \cdots & y_n^s(L-1)\end{bmatrix}^T$,  ${\cal D}(f_d) = \diag\left(1, e^{j2\pi f_d},\ldots,e^{j2\pi f_d(L-1)}\right)$, and ${\bf X}_{l_\tau}$ is the $l_\tau$ samples delayed version of transmitted signal matrix
\begin{eqnarray}
{\bf X} &=& \begin{bmatrix}
        x_1(0) & x_1(1) & \cdots & x_1(L-1) \\
        x_2(0) & x_2(1) & \cdots & x_2(L-1) \\
        \vdots & \vdots & \cdots & \vdots \\
        x_{N_T}(0) & x_{N_T}(1) & \cdots & x_{N_T}(L-1) \\
\end{bmatrix} \in {\cal C}^{N_T\times L}. \notag
\end{eqnarray}
The received signal in \eqref{eq:RxSig} serves as a basis for estimating the range, angular location, and velocity of the target. Relevant literature on parameter estimation can be found in \cite{xu2008target, Seifallah_Sajid_TSP}.

\subsection{Received Signal Model for Communication}
The transmitted signal from the DFRC system will also be received by multiple users. The received signal by $k$th user having a single antenna and flat-faded channel between the transmitter and user can be written as
\begin{eqnarray}
r^c_k(t) =  {\bf h}_k^T{\bf x}(t) + {\bf v}^c_k(t), \label{eq:RxSignalModel2}
\end{eqnarray}
where ${\bf h}_k = \begin{bmatrix} h^k_1 & h^k_2 & \cdots & h^k_{N_T}\end{bmatrix}^T$. $h^k_m$ is the channel gain between the $k$th user and the $m$th transmit antenna, and ${\bf v}^c_k(t)$ contains the contribution of noise and interfering signals from other targets. The signal model given in \eqref{eq:RxSignalModel2} can also be used for beamforming and transmitting information symbols. If we have $K$ users and the transmitter transmits $L$ symbols for each user, sampling \eqref{eq:RxSignalModel2}, the collection of received samples by all users in matrix form can be written as
\begin{eqnarray}
{\bf Y}^c =  {\bf H}_c{\bf X} + {\bf V}^c.  \label{eq:RxSigC}
\end{eqnarray}
Here, ${\bf Y}^c = \begin{bmatrix} {\bf y}^c_1 &  {\bf y}^c_2 & \cdots & {\bf y}^c_K \end{bmatrix}^T \in {\cal C}^{K\times L}$ while ${\bf y}^c_k =\begin{bmatrix} y^c_k(0) &  y^c_k(1) & \cdots & y^c_k(L-1)\end{bmatrix}^T$, ${\bf H}_c = \begin{bmatrix} {\bf h}_1 &  {\bf h}_2 & \cdots & {\bf h}_K \end{bmatrix}^T \in {\cal C}^{K\times N_T}$, and ${\bf V}^c \in {\cal C}^{K\times L}$ is the noise matrix.

The design of a DFRC system is mainly centered on two signal models as described in \eqref{eq:RxSig} and \eqref{eq:RxSigC}. 
In the following sections, we will delve into the latest advancements in DFRC system design.

\section{Performance Bounds for DFRC Systems}
\begin{figure}[htbp]
\centerline{\includegraphics[width = 0.5\textwidth]{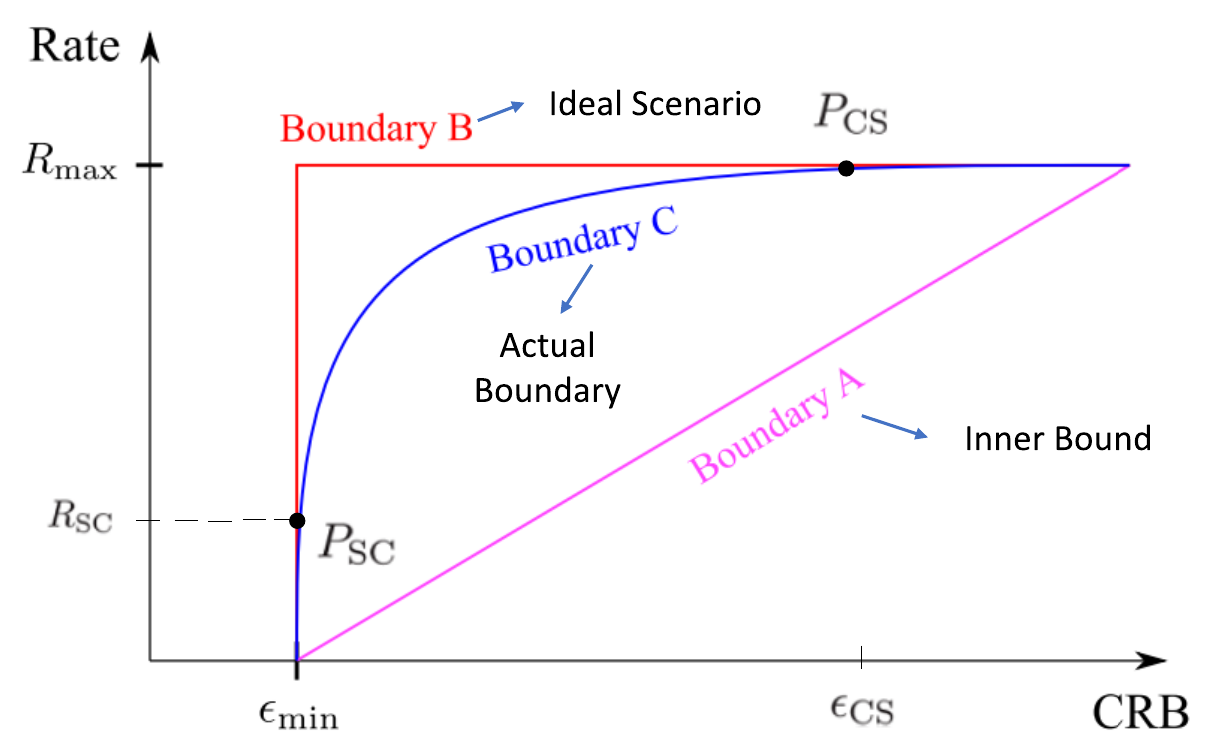}}
\caption{Various performance bounds in the CRB-rate region \cite{FanLiu_TIT2023}. Used under CC license.} \label{tradeoff}
\end{figure}

To gauge the performance of a DFRC system, it becomes pertinent to understand and analyze various performance bounds that can be achieved by a joint system as compared to independent communication and radar systems. In \cite{FanLiu_TIT2023}, it has been proposed to employ the sensing Cramér-Rao bound (CRB) and communication capacity metrics for investigating the fundamental limits of Sensing and Communication (S\&C) performance tradeoff. A Miller-Chang type CRB \cite{1055879} is employed for the analysis of the sensing performance. From Fig. \ref{tradeoff}, we can have three types of bounds/boundaries of the CRB-rate region namely
\begin{enumerate}
    \item Ideal Scenario and the Outer Bound: The ideal scenario is in which the DFRC system is able to achieve the maximum communication rate, $R_{max}$, along with minimum sensing CRB, $\epsilon_{min}$. In this case, there is no tradeoff between S\&C functionalities and both can achieve their optimum without degrading the other. This ideal boundary is depicted by the boundary B in Fig. \ref{tradeoff}. The outer bound will be closer to this ideal boundary and is obtained by letting the coherent sensing period go to infinity\cite{FanLiu_TIT2023}.
    \item Inner Bound: This is the lower bound of the CRB-rate region which can be achieved by using the time-sharing strategy \cite{el2011network} in which the sensing optimal strategy is applied with probability $p$ and the communication optimal strategy is applied with probability $1-p$. In this scenario, the integration of sensing and communication doesn't provide any additional performance gain. It is depicted by the boundary A in Fig. \ref{tradeoff}.
    \item Actual Boundary: This boundary depicts the actual performance of the DFRC system depending upon the tradeoffs involved. It is depicted by the boundary C in Fig. \ref{tradeoff}. Hence, the DFRC system will offer a performance that lies in between the ideal and inner boundaries and the aim is to get as close to the outer bound as possible by optimizing the tradeoffs involved. The S\&C tradeoff is characterized by the two corner points of the CRB-rate region. $P_{CS}$ indicates the minimum achievable CRB constrained by the maximum communication rate and $P_{SC}$ indicates the maximum achievable communication rate constrained by the minimum CRB.
\end{enumerate}

The tradeoff between sensing and communication is two-fold and includes the Subspace Tradeoff (ST) and the Deterministic-Random Tradeoff (DRT) \cite{FanLiu_TIT2023, fanliu-DRT} as discussed below. Depending on these tradeoffs, an outer bound and multiple inner bounds for the achievable CRB-rate regions have been proposed in \cite{FanLiu_TIT2023}.
\subsection{Subspace Tradeoff (ST)}
\begin{figure}[htbp]
\centerline{\includegraphics[width = 0.5\textwidth]{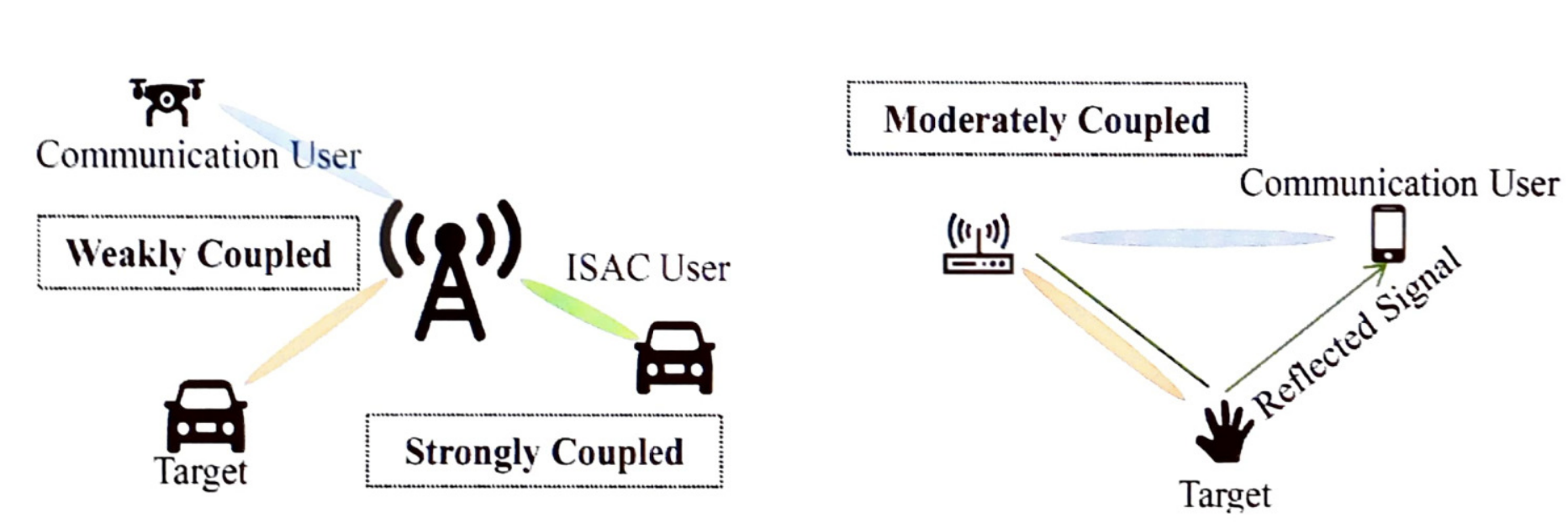}}
\caption{Coupling between sensing and communication channels \cite{boundf}.}\label{coup}
\end{figure}

The subspace tradeoff depends on the coupling strength between the subspaces spanned by S\&C channels thus balancing the resource allocation between sensing and communication subspaces. The greater is the coupling between two channels, the lesser the subspace tradeoff. The coupling strength can be strong, moderate, or weak as depicted in Fig. \ref{coup}. Strong coupling occurs between S\&C channels when the target to be sensed is also the communication user. Moderate coupling is when the sensing target is part of the communication channel and weak coupling is when both S\&C channels are different.  Mathematically speaking, given a specific statistical covariance matrix, $\widetilde{\boldsymbol{R}}=\mathbb{E}\left(\mathbf{R}\right)$, favorable communication performance is achieved when its column space is more closely aligned with the communication subspace, and vice versa.

\subsection{Deterministic-Random Tradeoff (DRT)}
It is a well-established fact that communication signals should be as random as possible to convey maximum information. Contrary to that, radar systems favor deterministic signals to achieve a stable detection performance. In \cite{FanLiu_TIT2023,fanliu-DRT}, the authors have explored the fundamental performance limits of a DFRC system from a rate-distortion perspective employing a vector Gaussian channel model. Conditional mutual information is used as a performance metric for both S\&C functionalities. The key findings indicate that sensing MI sets a universal lower limit for any well-defined distortion metrics for sensing and the distortion lower bound is achieved when the sample covariance matrix of the DFRC signal is deterministic. Hence, the DFRC system compromises signal randomness, thereby reducing communication performance, to enhance sensing performance. This is known as the deterministic-random tradeoff between sensing and communication performance in the DFRC system. The communication-optimal point, $P_{CS}$, can be achieved using conventional Gaussian signalling and the study has provided the lower and upper bounds for the sensing CRB, $\epsilon_{CS}$, at this point. The communication capacity, $R_{SC}$, for the sensing-optimal point, $P_{SC}$, can be achieved by a novel strategy based on uniform sampling over the set of semi-unitary matrices, i.e., the Stiefel manifold \cite{do1992riemannian}.


\section{Design of Dual Function Radar Communication Systems}
As far as the current literature is concerned, a vague boundary exists between the DFRC systems and the first two categories (JCR and JRC) of JCAS systems. However, there is more freedom of design in the case of DFRC, as we are no longer limited by existing radar and communication standards. We can design and optimize by considering the application-specific essential requirements for both functions, thus, leading to a better and more flexible trade-off between radar and communication performance. In this pretext, joint optimal waveform design becomes the fundamental problem of interest.

\begin{figure}[htbp]
\centerline{\includegraphics[width = 0.5\textwidth]{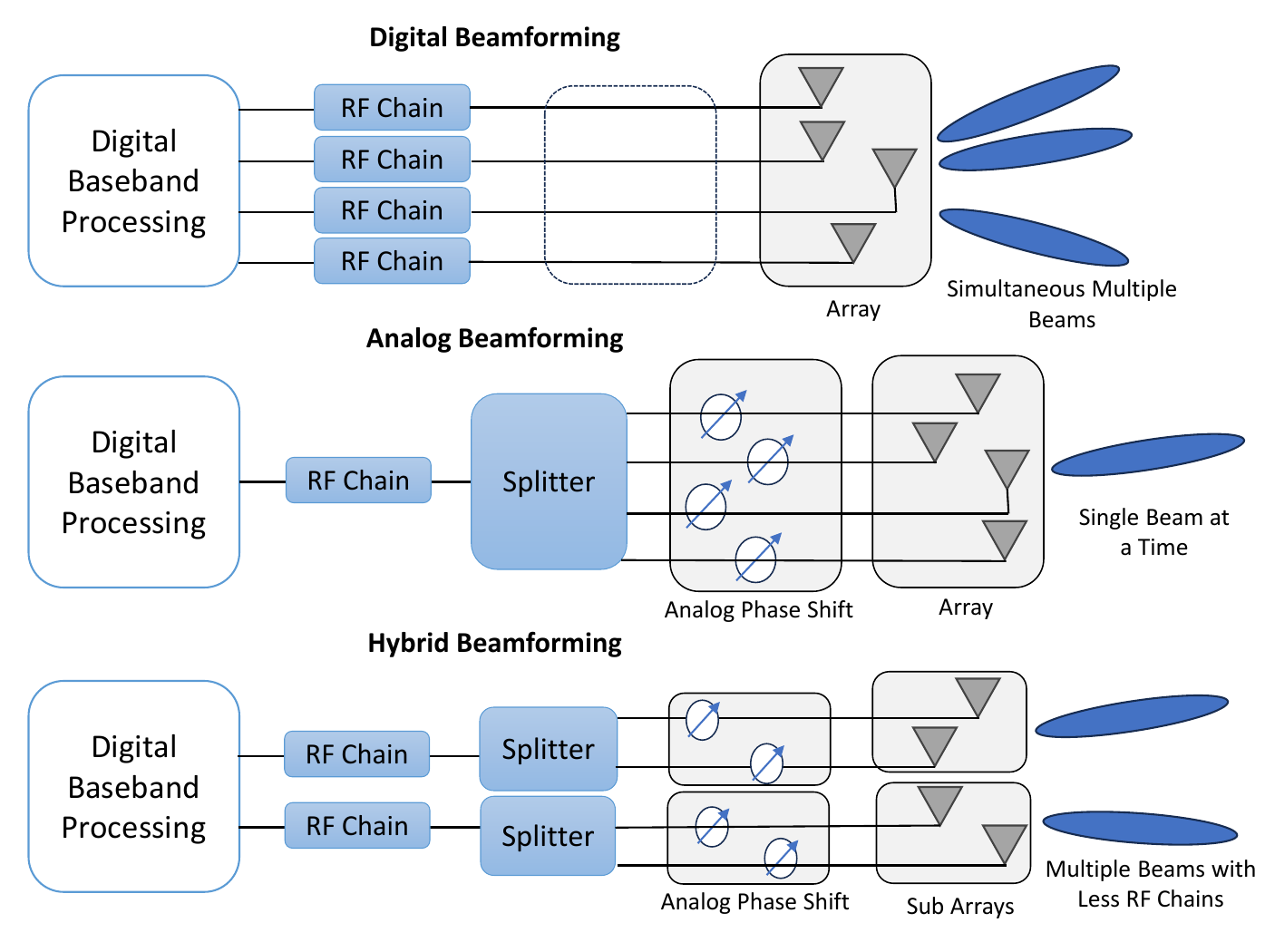}}
\caption{Digital, analog, and hybrid beamforming approaches for DFRC systems.}\label{beam}
\end{figure}

Hardware for DFRC systems can be categorized into phased array, fully digital Multi-Input Multi-Output (MIMO), and hybrid array architectures with corresponding beamforming approaches known as analog, digital, and hybrid analog-digital (HAD) beamforming. These approaches are depicted in Fig. \ref{beam}. In the case of analog beamforming, the same signal is fed to all the antennas through phase shifters, thus, steering a single beam in the desired direction. It is a simple, power-efficient, and cost-effective solution but can only generate a single beam and so suffers from beam squint effect for wideband transmission. For digital beamforming, each antenna is connected to a separate RF chain and signals are designed in the digital baseband using spatial precoding as done in MIMO systems. It has more degrees of freedom, can generate multiple beams in different directions and can support wideband transmission, however, it is complex, costly and requires more power. HAD is a compromise between the two approaches in which sub-arrays are formed from a large array and analog beamforming is used within each sub-array. It can offer the combined advantages of analog and digital BF while minimizing their drawbacks. Depending on the hardware architecture being employed for the DFRC system, we will now be discussing different waveform design and optimization approaches.
%
%

\subsection{Waveform Optimization for DFRC Systems with Digital Beamforming} \label{1}
In the case of digital beamforming for fully digital MIMO DFRC systems, the beampattern depends on the precoding matrix $\mathbf{W}$ such that $\mathbf{X=WS}$, where $\mathbf{S}$ is the desired constellation symbol matrix. Hence, the basic optimization problem can be formulated as \cite{hassanien2019dual}
\begin{equation}
\arg \max _{\mathbf{W}} f(\mathbf{W}), \quad \text { s.t. some constraints } \text {,}
\end{equation}
where $f(\mathbf{W})$ is the objective function. Different waveform optimization approaches are reported in the literature depending on how you define the objective function and the constraints. They can be defined individually for radar sensing or communications or in the form of a weighted joint function as discussed next.
\subsubsection{Mutual Information based Optimization}
Mutual Information (MI) is well-known for communication systems where the capacity of a channel is defined as the maximum mutual information between source and receiver. The first use of MI for radar waveform design for extended targets appears in \cite{259642}. MI for radars refers to the amount of information obtained about the channel from the signal received at the receiver. The conditional MI for radar and communication is defined as \cite{9303435,hassanien2019dual}
\begin{eqnarray}
I_S\left(\mathbf{H}_s ; \mathbf{Y}_s \mid \mathbf{X}\right) &=& M \log _2\left(\operatorname{det}\left(\frac{1}{\sigma_z^2} \mathbf{X}^H \boldsymbol{\Sigma}_\mathbf{H_s} \mathbf{X}+\mathbf{I}_L\right)\right), \notag \\
I_C\left(\mathbf{X} ; \mathbf{Y}_c \mid \mathbf{H}_c\right) &=& K \log _2\left(\operatorname{det}\left(\frac{1}{\sigma_z^2} \mathbf{H}_c^H \boldsymbol{\Sigma}_\mathbf{X} \mathbf{H}_c+\mathbf{I}_M\right)\right). \notag
\end{eqnarray}
Here, $\mathbf{H}_c$ and $\mathbf{H}_s$ are the communication and radar channels, respectively,   $\boldsymbol{\Sigma}_\mathbf{H_s}=\mathrm{E}\left[\mathbf{H}_{\mathbf{s}} \mathbf{H}_{\mathrm{s}}{ }^H\right] / M$, and $\boldsymbol{\Sigma}_\mathbf{X}=\mathrm{E}\left[\mathbf{X} \mathbf{X}^H\right] / L$. In \cite{9303435,7970102}, a weighted sum of these two MIs is used as an objective function for optimization
\begin{equation} \label{eq33}
F=\frac{w_R}{F_S} I_S\left(\mathbf{H}_s ; \mathbf{Y}_s \mid \mathbf{X}\right)+\frac{1-w_R}{F_C} I_C\left(\mathbf{X} ; \mathbf{Y}_c \mid \mathbf{H}_c\right),
\end{equation}
where $\mathbf{w}_R \in[0,1]$ is a weighting factor. $F_S$ and $F_C$ are the maximal MI for radar and communications. The objective function in (\ref{eq33}) can be optimized with the Karush-Kuhn-Tucker (KKT) conditions with an optimal water-ﬁlling type solution. 

\subsubsection{Beampattern Similarity based Optimization} \label{sec1}
Another approach relies on the optimization of the DFRC waveform with respect to a benchmark signal having a desired beampattern. The benchmark signal is usually a known radar waveform having desirable properties for sensing such as low sidelobe levels and good correlation. This is done in order to minimize the effect of the randomness of the communication data and channels on the waveform structure \cite{hassanien2019dual} and is achieved by using a waveform/beampattern similarity metric. Hence, the designed waveform has a beampattern that matches the desired radar's beampattern while satisfying the communication requirements.

In the study conducted by Liu et al. \cite{liu2018toward}, the authors propose closed-form dual-functional waveform designs for a MIMO system that serves both omnidirectional and directional radar beampatterns while minimizing the Multi-User Interference (MUI) for downlink communication users. For instance, to transmit an omnidirectional beam for sensing and symbol matrix ${\bf S}$ for communication users, the cost function is defined as
\begin{align}
\min_{\bf X} \|\mathbf{H}_c{\bf X}-\mathbf{S}\|_F^2, \notag \\
\text { s.t. } \frac{1}{L}\mathbf{XX}^H=\frac{P_T}{M}{\bf I}_M, \label{eq:OP1}
\end{align}
where $P_T$ is the total transmit power from all antennas. The authors derive a closed-form solution for this problem as
\begin{equation}
{\bf X} = \sqrt{\frac{LP_T}{M}}{\bf UI}_{M\times L}{\bf V}^H.
\end{equation}
Here, ${\bf U} \in {\cal C}^{M\times M}$ and ${\bf V}\in {\cal C}^{L\times L}$ are the singular-value-decomposition (SVD) matrices of ${\bf H}_c^H{\bf S}$ matrix. In the context of radar systems, efficient focusing of transmitted power toward specific targets is crucial. Therefore, for the desired beampattern that depends on the waveform covariance matrix, ${\bf R}_d$, Liu and his colleagues modify the original optimization problem as
\begin{align}
\min_{\bf X} |\mathbf{H}_c{\bf X}-\mathbf{S}\|_F^2, \notag \\
\text { s.t. } \frac{1}{L}\mathbf{XX}^H={\bf R}_d. \label{eq:OP2}
\end{align}
By employing the Cholesky decomposition of ${\bf R}_d = {\bf FF}^H$, the authors successfully derived a closed-form solution for the waveforms as
\begin{equation}
{\bf X} = \sqrt{L}{\bf F}\tilde{\bf U}{\bf I}_{M\times L}\tilde{\bf V}^H,
\end{equation}
where $\tilde{\bf U} \in {\cal C}^{M\times M}$ and $\tilde{\bf V}\in {\cal C}^{L\times L}$ are the SVD matrices of ${\bf F}^H{\bf H}_c^H{\bf S}$ matrix.

They also proposed weighted optimizations to create dual-functional waveforms that strike a flexible balance between radar and communication performance. The trade-off design becomes
\begin{equation} \label{eq1}
\begin{aligned}
& \min _{\mathbf{X}} \rho\|\mathbf{H_c X}-\mathbf{S}\|_F^2+(1-\rho)\left\|\mathbf{X}-\mathbf{X}_0\right\|_F^2, \\
& \text { s.t. } \frac{1}{L}\|\mathbf{X}\|_F^2=P_T (\mbox {total power constraint),}\notag \\
& \mbox{or} \notag\\
& \frac{1}{L} \operatorname{diag}\left(\mathbf{X} \mathbf{X}^H\right)=\frac{P_T}{M} \mathbf{1} \mbox{ (per antenna power constraint).}
\end{aligned}
\end{equation}
%
Here, $0\leq \rho \leq 1$ is a weighting factor that determines the performance weightage for radar and communication, and $\mathbf{X_0}$ is the reference radar waveform. Low complexity algorithms for the solution of (\ref{eq1}) are also presented. It was shown that a trade-off exists between the probability of detection and the achievable communication rate. For a fixed probability of detection, the achievable communication rate increases with the decrease in the number of users.

The constant modulus constraint is also desirable in practical radar systems where non-linear amplifiers typically work in a saturation condition, preventing amplitude modulation in radar waveforms \cite{6649991}. Waveform design that minimizes the communication MUI subject to the constant modulus constraint and similarity constraint, is also proposed in \cite{liu2018toward}. The optimization problem is formulated as \cite{liu2018toward}
\begin{equation}
\begin{gathered}
\min _{\mathbf{X}}\|\mathbf{H}^c{\bf X}-\mathbf{S}\|_F^2, \\
\text { s.t }\left\|\operatorname{vec}\left(\mathbf{X}-\mathbf{X}_0\right)\right\|_{\infty} \leq \eta, \\
\left|x_{i, j}\right|=\sqrt{\frac{P_T}{M}}, \forall i, j,
\end{gathered}
\end{equation}
where $\mathbf{X_0}$ is the reference radar signal matrix with constant modulus entries $x_{i,j}$. A branch-and-bound algorithm \cite{tuy2016convex} was developed to achieve a globally optimal solution, surpassing the performance of the conventional Successive Quadratic Constrained Quadratic Programming (QCQP) Refinement (SQR) algorithm \cite{aldayel2016successive}. Results of the constant modulus joint waveform design show that a trade-off exists between 1) communication sum-rate and radar waveform similarity tolerance (i.e. similarity between achieved and desired waveform), and 2) communication sum-rate and the radar pulse compression performance. Moreover, the omnidirectional beampattern design always outperforms the directional one in terms of communication sum-rate and symbol error rate.

In \cite{8288677}, the optimization problem aims to achieve a desired radar beampattern while ensuring that the SINR, $\gamma_k$, for $K$ single-antenna downlink users, remains above a certain threshold $\Gamma_k$
\begin{equation}
\begin{gathered}
\min _{\mathbf{W}, \beta}\left\|\mathbf{W} \mathbf{W}^H-\beta \mathbf{R}\right\|_F^2, \\
\text {s. t.} \beta \geq 0, \quad \gamma_k \geq \Gamma_k, \forall k ; \operatorname{diag}\left(\mathbf{W W}^H\right)=\frac{P_T}{M} \mathbf{1}_{M}.
\end{gathered}
\end{equation}
Here, $\beta$ is a scaling factor. $\mathbf{W}$ is the pre-coding matrix to be designed, and $\mathbf{R}$ is the covariance matrix of the radar waveform with the desired beampattern. This non-convex optimization problem is solved in \cite{8288677} by using efficient manifold algorithms to obtain a sub-optimal solution. Another interesting result reported in \cite{8288677} shows that the jointly designed waveform outperforms the case where communication and radar signals are statistically independent of each other.

To attain optimal performance, sensing systems necessitate the repetition of waveforms at predefined intervals, while communication systems demand waveform independence. In essence, for peak efficiency, the waveform covariance matrix must align for sensing and differ for communication systems \cite{FanLiu_TIT2023}. Previous studies, including the cited ones in this paper, have concentrated on waveform design to optimize both functions. However, these approaches often compromise one function to enhance the other. Recently, we introduced the concept of beampattern-invariant covariance matrix design, holding the promise of maximizing performance for both functions \cite{10205519}. This approach demonstrates that different covariance matrices can yield similar beampatterns as long as the sum of the diagonal elements in these matrices remains constant. Hence, the additional degrees of freedom can be exploited to convey the communication information. Mathematically, for a desired beampattern $P(\theta, \textbf{R})$ specified by the positive semidefinite and symmetric $M\times M$ covariance matrix {\bf R}, there exists a set $\mathcal{S}_1$ of covariance matrices defined as
\begin{equation}
\begin{aligned}
{\cal S}_1:=\bigg\{ & \tilde{\mathbf{R}}^{M \times M}: \sum_{k=1}^{M-n} \tilde{r}_{k, k+n}=\sum_{k=1}^{M-n} r_{k, k+n} \\
& \quad \text { for } n=0,1, \ldots M-1, \tilde{\mathbf{R}} \geq 0, \tilde{\mathbf{R}}=\tilde{\mathbf{R}}^H\bigg\}
\end{aligned}
\end{equation}
\textit{Then, for all} $\tilde{\mathbf{R}} \in \mathcal{S}_1$,
\begin{equation}
P(\theta, \tilde{\mathbf{R}})=P(\theta, \mathbf{R}).
\end{equation}
Hence, communication symbols can be transmitted by utilizing various covariance matrices without compromising the radar functionality. Moreover, an improved symbol error rate can be achieved by selecting the symbol set, from various possible symbol sets originating from different covariance matrices, that maximize the minimum distance among the symbols. 

\subsubsection{Cramér-Rao Bound (CRB) Based Optimization}
Accurate estimation of sensing parameters is essential in radar systems. To achieve optimal sensing, it's vital to minimize the mean-square estimation error (MSEE) for parameters like range, angular location, and velocity. Since Cram\'{e}r-Rao Lower Bound (CRLB) sets a minimum limit on the MSEE, it serves as a reliable performance metric for sensing. CRLB is influenced by the transmitted waveforms, so when designing waveforms for the DFRC system, it's crucial to utilize CRLB in the design process.
The CRLBs are well-established for radar signal-based estimates, as documented in \cite{6178073}. Furthermore, for communication signals, CRLBs for certain sensing parameters based on beamspace channel models are also been derived \cite{4770162}.

Due to the nonlinear relationship between the communication and sensing functions, it becomes challenging to optimize CRLB and mutual information of the communication system \cite{hassanien2019dual}. In \cite{liu2017multiobjective}, for instance, a single-antenna Orthogonal Frequency-Division Multiplexing (OFDM) DFRC system is considered to simultaneously improve the range, velocity accuracy, and the channel capacity for communications. However, due to the multi-objective nature of the optimization problem, the obtained solutions are not optimal. Another study, \cite{ni2020waveform}, examines waveform optimization by comparing and applying multiple sensing performance metrics, including MI, CRLB, and an approximated SINR metric for communications. The work in \cite{ni2020waveform} shows that significant correlations exist between MI-based and CRLB-based methods, highlighting that the MI-based approach is both more efficient and less complex compared to the CRLB-based approach.

The recent work in \cite{liu2021cramer} proposes MIMO beamforming designs for joint radar sensing and multi-user communications that minimize the CRLB for radar sensing while ensuring a predefined level of SINR for each communication user. For the single-user scenario, the authors derive a closed-form solution that optimizes the beamforming design for both point and extended targets. In the multi-user scenario, they demonstrate that the optimization problems can be relaxed by utilizing the semidefinite relaxation approach to obtain globally optimum solutions. Through simulation results, the authors demonstrate the superiority of the proposed method over beampattern approximation-based DFRC designs discussed in Section \ref{sec1}.

\subsection{Multibeam Optimization for DFRC Systems with Analog Beamforming} \label{2}
\begin{figure}[htbp]
	\centerline{\includegraphics[width = 0.5 \textwidth]{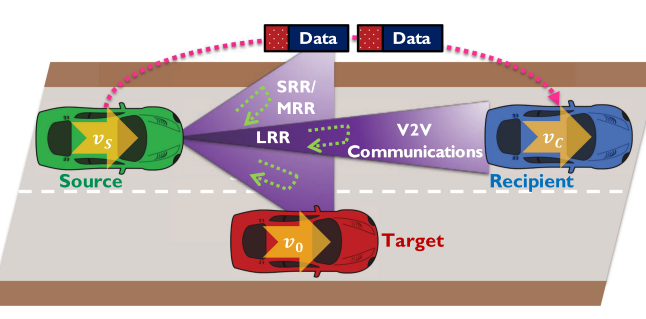}}
	\caption{Illustration of multibeam approach for vehicular networks \cite{9398548}. Used under CC license.}
	\label{beam1}
\end{figure}

In the context of mmWave JCAS systems, digital MIMO systems may not always be feasible due to their high hardware complexity and cost. Steerable analog or hybrid arrays are the alternative cost-effective options \cite{heath2016overview}. Thus, analog arrays, also the fundamental component of hybrid arrays, are faced with the challenge of supporting communication and sensing in different directions required by JCAS systems. To address this challenge with the limitations of analog array beamforming (BF) capability, a promising solution is the use of multibeam technology \cite{zhang2017joint,luo2019optimization,9131843}. The multibeam approach involves a fixed subbeam for communication and a scanning subbeam for sensing that varies its direction over different communication packets as shown in Fig. \ref{beam1}. Both subbeams possess identical information and are utilized for both communication and sensing.

Two main categories of methods have been proposed for multibeam optimization. The first category is the subbeam combination method, as described in \cite{zhang2017joint,luo2019optimization}. To generate the fixed and scanning subbeams, two beamforming vectors, $\mathbf{w}_{t, c}$ and $\mathbf{w}_{t, s}$, are designed respectively. These vectors are combined using a phase shifting coefficient and a power distribution factor to form the BF vector $\mathbf{w}_{T}$ as
\begin{equation} \label{eq4}
\mathbf{w}_T=\sqrt{\rho} \mathbf{w}_{t, c}+\sqrt{1-\rho} e^{j \varphi} \mathbf{w}_{t, s},
\end{equation}
where $\rho  (0 < \rho < 1)$ is the power distribution factor and $e^{j \varphi}$ is the phase shifting coefficient. Balancing the communication and sensing distances typically determines the value of $\rho$. The optimization process primarily revolves around $\varphi$, as it exerts a significant influence on the performance of the BF system. When designing the BF vectors for the subbeams ($\mathbf{w}_{t, c}$ and $\mathbf{w}_{t, s}$), the optimization primarily focuses on the magnitudes of the desired BF waveform, neglecting the consideration of phases. The effectiveness of the multibeam is dependent on how two BF vectors are combined. Through the optimization of $\varphi$, the coherency between the pre-generated subbeams is ensured, enabling the formation of a coherent multibeam.

One optimization problem, using the subbeam method, that maximizes the SNR for communications with constrained beamforming gain on the sensing subbeams, is formulated as \cite{9131843}
\begin{equation}
\begin{aligned}
& \varphi_{\mathrm{opt}}=\arg \max _{\varphi} \frac{\mathbf{w}_T^H \mathbf{H}_c^H \mathbf{H}_c \mathbf{w}_T}{\left\|\mathbf{w}_T\right\|^2}, \\
& \text { s.t. } \frac{\left|\mathbf{a}_T^T\left(M, \phi_i\right) \mathbf{w}_T\right|^2}{\left\|\mathbf{w}_T\right\|^2} \geq C_i^2(1-\rho) M, i=1,2, \ldots, N_s.
\end{aligned}
\end{equation}
Here, $\mathbf{w}_{T}$ is the BF vector in (\ref{eq4}), $C_i$ is the ratio of the scanning subbeam gain to the maximum possible array gain, i.e., $(1 - \rho)M$, $\varphi_i$s are the desired scanning directions, and $N_s$ is the total number of constraints. Closed-form solutions for such problems are derived in \cite{9131843}.

While the subbeam combination method is simple and efficient for implementation, it is considered suboptimal due to the separate pre-generation of BF weights for the two subbeams, combined with only a single variable. The extent of its performance gap compared to the optimum, and the existence of the optimum solution remains unclear. To address this issue, global optimization methods are developed in \cite{9131843} to directly optimize the BF vector  $\mathbf{w}_{T}$, taking into account the requirements of both communication and sensing. An example of such global optimization can be as follows, where the BF vector is optimized to maximize the received signal power for communication subject to similarity and gain constraints \cite{9131843}
\begin{equation} \label{eq5}
\begin{array} {c}
\mathbf{w}_{t, \text { opt }}=\operatorname{argmax}_{\mathbf{w}_T, \mathbf{w}_T^H \mathbf{w}_T=1} \mathbf{w}_T^H \mathbf{H}_c^H \mathbf{H}_c \mathbf{w}_T,\\
\\
\text { s.t. }  \left.\| \mathbf{A}\left(M, \phi_i\right) \mathbf{w}_T-\mathbf{d}_v\right) \|^2 \leq \varepsilon_w, \text { and/or } \\
\quad\left|\mathbf{a}_T^T\left(M, \phi_i\right)^T \mathbf{w}_T\right|^2 \geq \varepsilon_i, i=1,2, \ldots, N_s.
\end{array}
\end{equation}
The first constraint is the similarity constraint between the generated  BF waveform and the desired $\mathbf{d}_v$, the second is the BF gain constraint in the scanning directions, and $\varepsilon_w$, $\varepsilon_i$ are the respective thresholds. The optimization problem in (\ref{eq5}) is a non-convex NP-hard problem, which is solved in \cite{9131843} using the semidefinite relaxation approach after converting it to a homogeneous QCQP. Comparison of the global optimization method and the subbeam combination method (\ref{eq4}) highlights the superiority of the former in terms of achieved received signal power and the MSE of BF waveform. More recent works on multibeam optimization include \cite{9786835,10153585}.

Recent work in \cite{10038802} introduces energy-efficient and easy-to-design Multi-Beam Antenna Arrays for DFRC systems. These arrays utilize pre-designed analog devices, such as lenses or Butler matrix \cite{guo2021circuit}, to steer multiple beams with minimal power consumption. They offer flexible beam synthesis, accurate angle-of-arrival estimation, and convenient handling of the beam squint effect \cite{10038802}.

\subsection{Waveform Optimization for DFRC Systems with Hybrid Analog-Digital (HAD) Beamforming} \label{3}
\begin{figure*}[t]
	\centerline{\includegraphics[width = 0.95 \textwidth]{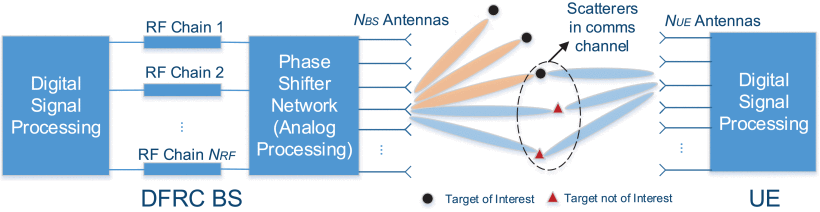}}
	\caption{mmWave massive MIMO DFRC system model with HAD beamforming used in \cite{liu2020joint}. \textcopyright 2020 IEEE.}
	\label{sys}
\end{figure*}

HAD represents a middle ground between digital and analog beamforming, creating sub-arrays from a larger array and employing analog beamforming within each sub-array as depicted in Fig. \ref{beam}. This approach aims to harness the benefits of both analog and digital beamforming while mitigating their respective limitations. In \cite{liu2020joint}, the authors proposed a novel transceiver architecture for a mmWave massive MIMO DFRC system employing HAD beamforming. A novel Time-Division Duplex (TDD) frame structure and signal processing approach was utilized to combine similar radar and communication operations in three stages of 1) radar target search and communication channel estimation, 2) downlink communication and radar transmit beamforming, and 3) uplink communication and radar target tracking. A joint receiver design is also proposed, capable of target tracking while decoding the uplink communication signals. The system model used in \cite{liu2020joint} is shown in Fig. \ref{sys}.

\begin{figure*}[htbp]
	\centerline{\includegraphics[width = 0.95 \textwidth]{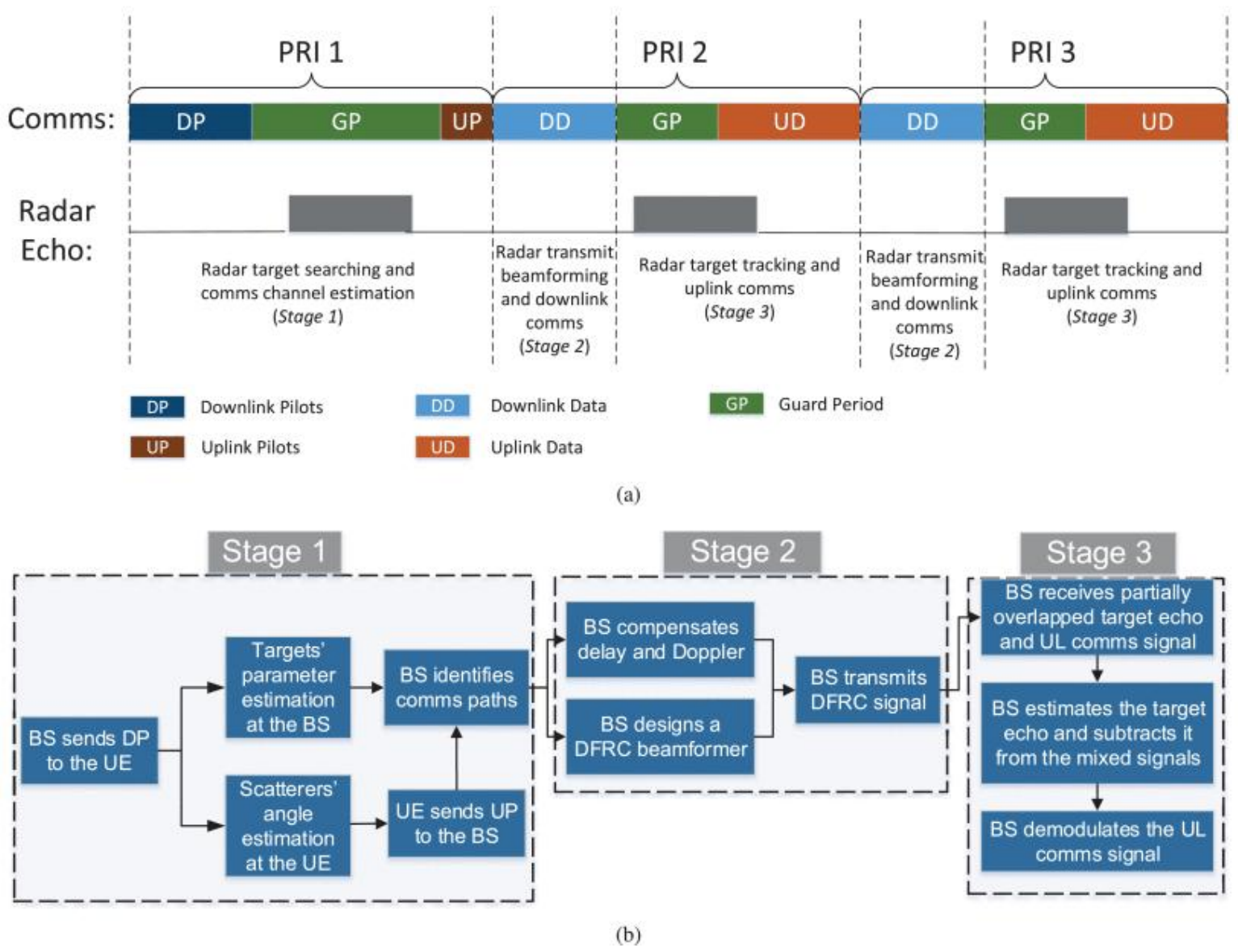}}
	\caption{ (a) Proposed frame structure and, (b) Signal processing flowchart used in \cite{liu2020joint}. \textcopyright 2020 IEEE.}
	\label{stage}
\end{figure*}

The frame structure and signal processing flowchart, proposed in \cite{liu2020joint}, are depicted in Fig. \ref{stage}. In the first stage, orthogonal Linear Frequency Modulation (LFM) pilot signals are generated from the HAD array for joint channel estimation and target search. Angle, delay and Doppler, and reflection coefficient are estimated at the base station (BS) using MUSIC \cite{schmidt1986multiple}, matched filtering \cite{roberts2010iterative}, and Angle and Phase EStimation (APES) \cite{li1996adaptive,xu2008target} algorithms, respectively. In the second stage, analog and digital beamforming matrices are designed to generate direction beams toward the targets and scatterers while equalizing the communication channel. In the last stage, target echoes and uplink data from the user are received at the BS. Target tracking is performed through peak search in the MUSIC spectrum within $\left[\hat{\theta}_k-\Delta_{\max }, \hat{\theta}_k+\Delta_{\max }\right]$, where $\hat{\theta}_k$ represents the estimated angles of the targets from the first stage and $\Delta_{\max }$ is the maximum angular variation. Based on the updated estimates of the reflection coefficients and angles of targets, the uplink communication signal is recovered by using the Successive Interference Cancellation (SIC) approach \cite{tse2005fundamentals}.

\subsection{Waveform Optimization across Time Domain}
Different works have also considered the waveform optimization problem across time and frequency domains apart from the spatial domain approach covered in sub-sections \ref{1} to \ref{3}. Sensing parameters' estimation accuracy of communication signals can be improved with waveform optimization in the time domain. A typical communication frame consists of a preamble followed by a data payload. In \cite{8917703}, the preamble of a communication frame is used for sensing, and sparse sensing in the time domain is exploited for the waveform design. It is shown that non-uniformly placed preambles can achieve better velocity estimation accuracy as compared to the uniform ones at the cost of a small drop in communication rate. In \cite{9398548}, a mmWave combined waveform-beamforming design is proposed for automotive applications with increased field-of-view. An optimized convolutional compressed sensing technique is developed that applies carefully designed circulant shifts of the transmit beamformer at the transmitter for superior radar channel reconstruction in the Doppler-angle domain using compressed sensing in the space-time domain. Performance trade-off between radar and communication is evaluated using a Normalized Mean Square Error metric for radar estimation and a Distortion MSE metric for communication. In \cite{9086030}, Doppler-resilient Golay complementary sequences across multiple packets, are constructed to achieve a 20 dB improvement in range sidelobe suppression over the standard 802.11ad protocol along with improvement in detection performance for a given SNR.

\subsection{Coherent Receiver Design for DFRC Systems}
In a typical radar system, same waveform is transmitted after each Pulse Repetition Interval (PRI) and multiple echoes are combined within a Coherent Processing Interval (CPI) to improve SNR, estimate Doppler, and suppress clutter \cite{richards2010principles}. However, the situation changes for the case of a DFRC system where a single waveform is employed for communication and radar sensing, and random communication symbols are also transmitted via the same waveform. Owing to the transmission of different symbol waveforms within a CPI, the output of the DFRC receiver's radar unit no longer remains coherent. This non-coherency can make Doppler estimation and clutter suppression really challenging and can also lead to weaker targets being masked by the range side lobes \cite{9905501}.

In \cite{5592502}, an iterative adaptation of the least squares mismatch filter approach is proposed to achieve coherent output response for two waveforms. However, the coherency is not guaranteed for more than two waveforms. A cascaded filter design is proposed in \cite{7944446} to achieve coherency at the expense of increased range side lobes. In \cite{9905501}, a novel fully coherent closed-form receiver design for an arbitrary number of waveforms is proposed based on linear and circular convolution, without degrading the performance of the DFRC system. Within each PRI (Pulse Repetition Interval), this system enables the transmission of any of $K$ distinct symbols. The suggested approach involves the creation of $K$ specific matched filters for the various symbols being transmitted. The challenge is framed as an optimization problem where each filter is designed to produce an impulse-like response when the corresponding symbol waveform is inputted while meeting coherent output response constraints for each filter. This approach maintains the integrity of both communication and sensing functions without any compromise.

\subsection{Physical Layer Security of DFRC Systems}
Wireless communication systems face inherent security vulnerabilities due to their broadcast nature. The integration of sensing and communication exacerbates this vulnerability, hindering the full potential of DFRC systems. In joint communication and radar sensing scenario, the targets to be detected could be potential eavesdroppers, especially for defense-related applications. For radar sensing, the DFRC BS must use high-power signals to gather information like angle and velocity from target echoes while suppressing side lobes to minimize clutter. To enable effective joint sensing and communication, the DFRC BS designs information-carrying signals to create highly directional sensing beams. However, this exposes the communication information to potential eavesdropping by malicious targets, making security a critical consideration in DFRC system design \cite{xu2022robust,su2019enhancing}. Artificial Noise (AN) has been widely exploited in the literature to enhance communication secrecy \cite{negi2005secret,su2019enhancing,xu2022robust}.

In \cite{su2019enhancing}, the authors proposed to enhance the physical layer security of the DFRC system by formulating fractional programming problems to minimize the SINR at the eavesdropper with the help of AN, while maintaining a desired SINR threshold for legitimate user. The downlink cellular users are considered as legitimate users while the radar targets are regarded as potential eavesdroppers. By adding artificial noise $\mathbf{N}$ to the transmitted signal, the transmit matrix $\mathbf{X}$ from a dual-functional MIMO BS can be written as
\begin{equation}
\mathbf{X}=\mathbf{W S}+\mathbf{N},
\end{equation}
where $\mathbf{S}$ is the desired signal from the BS and $\mathbf{W}$ is the beamforming matrix. The optimization problem formulated in \cite{su2019enhancing} for minimizing the SINR at the target is given as
\begin{equation}
\begin{array}{ll}
\min _{\mathbf{W}_i, \mathbf{R}_N} \frac{|\alpha|^2 \mathbf{a}^H\left(\theta_0\right) \sum_{i=1}^K \mathbf{W}_i \mathbf{a}\left(\theta_0\right)}{|\alpha|^2 \mathbf{a}^H\left(\theta_0\right) \mathbf{R}_N \mathbf{a}\left(\theta_0\right)+\sigma^2}, \forall i \\
\text { s.t. } \quad\left\|\mathbf{R}_X-\mathbf{R}_d\right\|^2 \leqslant \gamma_{b p}, \\
\quad \operatorname{SINR}_i \geqslant \gamma_b, \forall i, \\
\quad P_t=P_0.
\end{array}
\end{equation}
Here, $\theta_0$ represents the location of targets, $\gamma_{b p}$ is the threshold that constraints the mismatch between the designed covariance matrix $\mathbf{R}_X$ and the desired $\mathbf{R}_d$, $P_0$ is the transmission power budget, $\mathbf{R}_N$ is the covariance matrix of artificial noise, and $\gamma_{b}$ is the required SINR threshold for each legitimate user. By minimizing the SINR at the target, desired secrecy rate can be achieved.

A more recent work \cite{xu2022robust} proposes an optimization framework for robust and secure resource allocation for a DFRC BS by employing variable-length snapshots of the environment by scanning a pre-defined sector with a sequence  of dedicated beams. By jointly optimizing the duration of each snapshot, the beamforming vector, and the covariance matrix of the AN, the system sum secrecy rate $\sum_{k \in \mathcal{K}} R_k^{\sec }[m]$ in the $m$-th snapshot is maximized where $R_k^{\mathrm{sec}}[m]$ is the maximum achievable secrecy rate between the BS and the $k$-th legitimate user given by
\begin{equation}
R_k^{\mathrm{sec}}[m]=\left[R_k[m]-\max _{j \in \mathcal{J}} C_{j, k}[m]\right]^{+}.
\end{equation}
Here, $R_k[m]$ is the achievable rate of legitimate user $k$ in snapshot $m$ and $C_{j, k}[m]$ is the capacity of the channel between the DFRC BS and potential eavesdropper $j$ in snapshot $m$. Simulation results reported in \cite{xu2022robust} indicate that the proposed design can significantly enhance the physical layer security  of the DFRC system compared to other schemes that utilize single snapshot design.

\section{APPLICATION SCENARIOS FOR JOINT COMMUNICATION AND SENSING}
In this section, we will present several key application scenarios where joint communication and sensing systems demonstrate their utility beyond addressing spectrum congestion problem.

\subsection{Perceptive Mobile Networks}
Present wireless localization methods predominantly adopt a device-centric approach. In this paradigm, a wireless device, such as a smartphone, is affixed to the target object. The device's location is then deduced by analyzing signal interactions and geometric correlations with other stationed wireless equipment, such as a base station. Nevertheless, this device-oriented strategy restricts the range of applicable targets and lacks adaptability across varied scenarios. By capitalizing on supplementary Doppler processing and harnessing valuable information from multipath components,  JCAS-empowered cellular networks can enhance localization accuracy in contrast to prevailing localization technologies \cite{Cui2023}. A framework for a Perceptive Mobile Network (PMN) was first presented in \cite{rahman2019framework}. The idea is to integrate radar sensing into the current cellular networks by incorporating few modifications so that it has the ability to perceive the environment via radio vision and inference. In particular, a Cloud Radio Access Network (CRAN) architecture employing Spatial Division Multiple Access and OFDMA is considered in \cite{rahman2019framework}. Another possible architecture proposed for PMNs is utilizing Target Monitoring Terminals (TMTs) \cite{xie2022perceptive}. Direct and indirect sensing schemes based on 1-D compressive sensing were developed in \cite{rahman2019framework} for sensing parameters estimation. For direct sensing, the received signal is fed directly to the sensing algorithm, whereas signal stripping is used in indirect estimation to simplify the signal by removal of data symbols after demodulation and decorrelation of users using channel estimation. In addition, a background subtraction method for the suppression of clutter was also proposed.
\begin{figure}[htbp]
	\centerline{\includegraphics[width = 0.5 \textwidth]{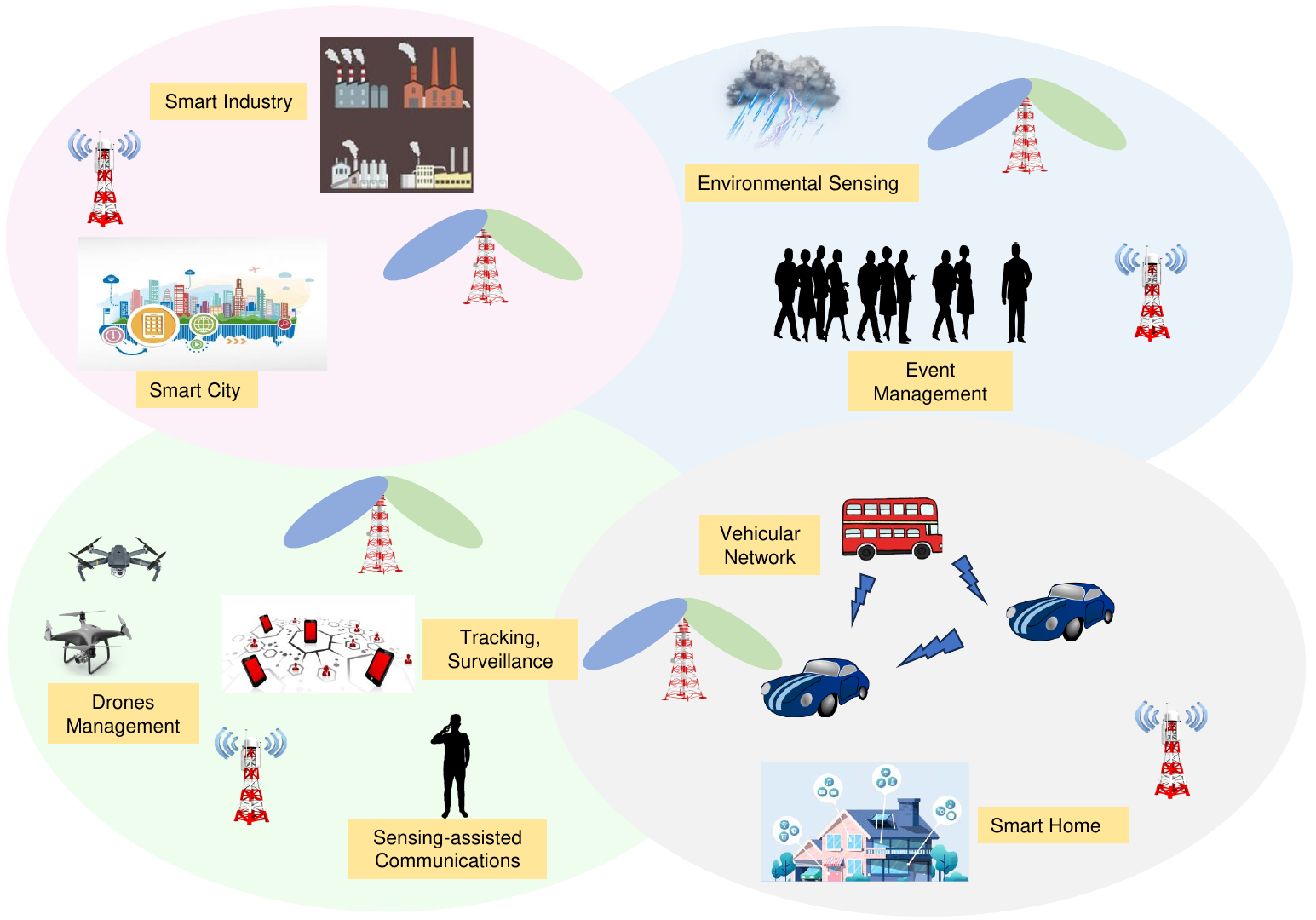}}
	\caption{Potential applications of PMNs with joint communication and sensing capabilities.} 	\label{pmn}
\end{figure}

PMNs can provide unobtrusive and ubiquitous sensing capability (which otherwise is either impractical or too costly to implement) because of their large coverage and infrastructure. In addition, they can provide networked sensing, where the perspectives from multiple sensing nodes can be collaboratively utilized for sensing the same target. PMNs can enable numerous new applications related to smart cities, smart homes, IoT, vehicular networks, smart industry, environmental sensing, drone monitoring and management, event management, passive sensing, surveillance, and sensing-assisted communications as depicted in Fig. \ref{pmn}. A detailed survey of the latest research on PMNs is presented in \cite{9585321,10077114}.

\subsection{Autonomous Vehicles}
Next generation of autonomous vehicles requires low-latency Gbps data rates and high-resolution sensing information on the order of few centimeters \cite{8246850}. Vehicular localization utilizes technologies like Global Navigation Satellite-based Systems (GNSS), radar, camera, and Light Detection and Ranging (Lidar). GNSS alone can't fulfill the localization requirements of the autonomous vehicles due to its low-resolution, high latency, and coverage holes \cite{8246850}. GNSS is assisted by dedicated base stations and the on-board sensors mentioned above but these sensors add to the cost and computational complexity. Vehicular communication (V2V, V2I, and V2X) uses standards such as Dedicated Short-Range Communication (DSRC) \cite{kenney2011dedicated} and the D2D mode of LTE-A. However, these standards are not able to support the demanding Gbps requirements of modern autonomous vehicles \cite{choi2016millimeter}. In this context, it is argued that 5G mmWave communication is the only viable solution for vehicular positioning and  data exchange between connected vehicles \cite{choi2016millimeter,8246850}. One key challenge to the mmWave vehicular communication is the overhead associated with beam training. In  \cite{choi2016millimeter}, it has been proposed to utilize the sensing information from the onboard sensors or DSRC as side information to the mmwave communication link in order to reduce the beam training overhead. PMNs can address this challenge effectively as they offer both the functionalities of communication and sensing. DFRC systems can be employed onboard the autonomous vehicles to offer both the functionalities of C\&R sensing providing a low power, reduced size, and low-cost solution. Use of DFRC systems in autonomous vehicles can also address possible interference of mmWave communication with automotive radars which also operate in the mmWave band. It becomes apparent from the discussion above that mmWave DFRC systems appear to be a promising candidate for modern autonomous vehicles.

\subsection{Drones}
There is a growing interest in utilizing UAVs as cost-efficient aerial platforms to deliver UAV-enabled communication services \cite{7470933} as exemplified by initiatives like AT\&T's flying COW (Cell on Wings) \cite{cow}. These services include supporting search and rescue efforts, extending coverage to remote areas, and enhancing services in temporary hotspot areas like concerts, football games,...etc. However, stringent requirements on SWaP (size, weight, and power) impose limitations on both fixed-wing and rotary-wing drones, constraining their communication, sensing, and endurance capabilities. The use of DFRC systems can significantly help in addressing the SWaP constraints on drones while fulfilling the communication and sensing requirements. It can also offer all-weather day/night sensing capabilities in contrast to the conventional camera sensor utilized by drones. Moreover, radar sensing capability can be exploited by drone swarms for formation flight and collision avoidance. Delivery drones, used for the delivery of small packages, can also benefit from the advantages offered by JCAS systems. From a military point-of-view, the employment of a joint system on drones can significantly help in search and rescue, surveillance, and electronic warfare operations by reducing the payload and radar cross-section of the drone, thus, improving its mobility and covertness.

A comprehensive overview of UAV-enabled joint communication and sensing is presented in \cite{10098686} where authors covered single and multi-UAV-enabled JCAS. Different application scenarios, emphasizing the benefits of UAV-enabled JCAS, resulting from the coordination between communication and sensing functionalities have been discussed under two categories of sensing-assisted UAV communication and communication-assisted UAV sensing. It is pertinent to note that, unlike conventional ground-based JCAS systems, in UAV-enabled JCAS systems, the allocation of resources and the design of waveforms are significantly impacted by the deployment and trajectory of the UAV. This is due to the changing angular separations between users and targets as the UAV's position changes. Therefore, achieving high-performance sensing and communication requires a joint design approach that encompasses user association, transmit beamforming, and route planning \cite{meng2022throughput}.

\subsection{Security and Surveillance}
JCAS systems can be a good potential candidate for perimeter security and surveillance of border areas and sensitive installations like military bases, airports, dams, and grid stations. Radar sensing can offer all-weather day/night surveillance as opposed to other sensors like cameras. The information collected by radar sensing can be simultaneously transmitted by the communication functionality of the JCAS system to the central control center for initiating countermeasures like the deployment of quick response forces. This will eradicate the need for installing separate sensor and communication equipment and laying physical media between various security posts/towers and the central control center, thus, providing an efficient and cost-effective solution. As a step further, micro-Doppler signatures can be extracted and analyzed for activity recognition and classification of various types of targets for better threat management \cite{9673798}.

\subsection{Smart Homes and Smart Industry}
With the rapid pace of technological advancement, concepts like smart home and smart industry, are becoming a reality. The potential of RF sensing in smart homes and industries has been well-established by research \cite{teixeira2010survey,adib2015smart,zou2017wi,lubna2022radio}. However, it contributes to network congestion due to the substantial amount of wireless traffic in both scenarios. Rather than designing communication and sensing systems in isolation, joint design can alleviate the strain on wireless resources. The issue can be addressed from two perspectives 1) to equip the conventional sensing device with communication functionality and, 2) to enhance the sensing prowess of WiFi signals by redesigning the packet format \cite{9143269}.

The integration of joint communication and sensing systems holds immense promise for a variety of applications within the realm of smart homes. The localization performance of conventional GPS degrades significantly in indoor scenarios. Wifi-based localization has emerged as a promising solution for indoor localization due to its low cost, ubiquitous deployment, and no additional hardware requirement \cite{9855454}. Through-wall localization and tracking enables accurate positioning of individuals even behind physical barriers and provide a foundation for responsive home automation. Existing Wifi-based localization either uses propagation-based or fingerprinting techniques \cite{8767421,9745151}. However, propagation-based techniques have low accuracy and they localize devices instead of users, whereas fingerprinting techniques are sensitive to human location, body shape, and the ambient environment \cite{9745151}.

JCAS systems can overcome these limitations by embedding sensing functionalities into the current Wifi systems. These systems can be deployed in a monostatic or a multistatic configuration like \cite{hex}. As a result, human-centric functionalities in smart homes can be vastly enhanced through these systems. Micro-Doppler signatures can be extracted for human activity recognition, health monitoring, and fall detection to ensure the safety and well-being of occupants \cite{chen2016activity,10168871,imamura2022automatic}. This real-time health insight empowers remote health management and facilitates early intervention. Gesture recognition can be employed for the operation and control of smart devices \cite{wan2014gesture}. The sensing functionality can also be exploited for intrusion detection and security \cite{hex}.

In the same context, joint communication and sensing can be used within the domain of smart industry for localization and tracking of vehicles, equipment, and workers \cite{9488464,cupek2020autonomous}; object recognition and authentication \cite{8355491}; surveillance of designated areas; and gesture recognition for equipment operation \cite{9585321}. By seamlessly integrating communication and sensing technologies, industries can achieve real-time monitoring and precise positioning, leading to enhanced operational efficiency, streamlined logistics, and improved safety protocols.

\subsection{Replacement/Redundancy of Backhaul Connectivity for Military Radars and Airborne Warning and Control System (AWACS)}
Modern military radars are usually networked together to provide a comprehensive picture of a country's airspace at the central command and control center. This greatly helps in effective decision making and operational command and control during war-time scenarios. Ground-based radars usually send their data to the central command and control center via dedicated optical fiber connectivity whereas AWACS share their radar picture via dedicated air-to-ground data links. By employing the communication functionality of radar-centric JCAS systems, this dedicated back-haul connectivity with the central command and control center is no longer required (especially in the case of remote and hilly areas), thus saving huge costs. Even if this back-haul connectivity is not fully replaced, the communication functionality of JCAS system can act as a backup for this back-haul connectivity. For the case of AWACS, it can offer another very critical advantage of weight reduction (resulting from the deletion/reduction of dedicated data link communication equipment), which leads to enhanced endurance for the platform.

\subsection{Multipurpose RF Systems for Military}
There are various RF systems onboard military ships and aircrafts serving different purposes of communication, radar, and electronic warfare. The independent design and development of these systems have led to a significant increase in the size, weight, and power requirements of the combat platform, leading to larger radar cross-section and lesser endurance for the platform. These sophisticated RF systems come with a hefty price tag and require costly maintenance and complex troubleshooting, resulting in increased downtime of the combat platform. Additionally, the coexistence of such systems inevitably causes mutual interference among each other.

In 1996, the Advanced Multi-function Radio Frequency Concept project was launched by the Defense Advanced Research Projects Agency (DARPA), to design integrated RF systems utilizing a common set of array architecture, signal and data processing, signal generation, and display hardware \cite{1406306}.  In 2013, Shared Spectrum Access for Radar and Communications project \cite{ssparc} was launched by DARPA. The program aims to improve radar and communications capabilities through spectrum sharing. In particular, two types of spectrum sharing were considered for the S-band (2-4 GHz):  spectrum sharing between military radars and military communications systems and spectrum sharing between military radars and commercial communications systems. An improved version of the program named Joint Design and Operation of SSPARC was proposed in \cite{7131098}. Clearly, all these projects aim for the fusion of radar and communication functionalities and DFRC systems possess the abilities to achieve these desired outcomes in an efficient way.

\section{RESOURCE ALLOCATION IN DFRC SYSYTEMS}
To achieve optimal Quality of Service (QoS) in DFRC systems, a deep understanding of radar and communication systems and individual system’s unique requirements is essential. Optimizing QoS in DFRC systems involves the strategic allocation of resources to strike a harmonious balance between the needs of sensing and communication tasks while adapting to changing requirements. Hence, effective resource management is fundamental to the seamless operation of these integrated systems across various operational scenarios.

\subsection{Memory Management}
Radar systems often necessitate substantial memory resources to store critical data utilized in target localization and tracking. This data includes both fast-time and slow-time samples, as well as historical information for tracking purposes. In the context of identifying slow-moving targets, the use of fast-time samples aids in estimating the target's range, while slow-time samples are pivotal for velocity estimation. The combined storage of fast and slow-time sample data can impose a considerable memory burden. Nonetheless, an efficient approach can be adopted to minimize memory utilization. The process begins with the application of a Fast Fourier Transform (FFT) to the fast-time samples, which facilitates the estimation of target ranges. The indices corresponding to spectral peaks obtained through this FFT process are then leveraged to estimate the ranges of different targets within the radar's field. Significantly, slow-moving targets exhibit minimal variations in their ranges over the course of a single CPI. This insight allows us to optimize memory usage. Specifically, after the initial pulse, the estimated range indices remain relatively consistent for the duration of a frame. Consequently, it becomes possible to employ the indices derived after the first pulse for subsequent pulses during the same frame. This strategic decision means that only samples associated with the range indices following the transmission of each pulse need to be preserved for velocity estimation. By adopting this technique, a substantial reduction in the required memory capacity can be achieved, without compromising the radar system's ability to perform accurate target localization and tracking.

In the case of radar systems, the emphasis is on efficiently storing and reusing range indices to minimize memory consumption while ensuring precise target localization and tracking. In contrast, communication units operate on a frame-by-frame data processing paradigm, which demands a distinct memory usage pattern compared to radar systems. A desired approach to memory resource management in DFRC systems ensures that both radar and communication units can effectively fulfill their respective functions. Therefore, the allocation of memory resources must be tailored to accommodate the unique requirements of each task.

In \cite{9413979}, the design of a communication receiver, operating under bit constraints, is considered for a radar-centric JCAS scenario. Task-based quantization framework \cite{shlezinger2019hardware,8736805,salamatian2019task,neuhaus2021task} is utilized to facilitate the operation of communication receiver under bit constraints. The proposed hybrid/analog receiver design employs a linear analog combiner before the quantization and digital processing stages. Two different types of analog combiners are proposed; the spatial-temporal combiner and the spatial combiner, which are jointly designed with the overall system. The analog combiner reduces the dimensionality of the input to the analog-to-digital converters while preserving the dominant eigenmodes of the linear minimum mean-square error estimate of the transmit symbol vector. The findings presented in \cite{9413979} illustrate that the proposed communication receiver yielded superior performance compared to the task-ignorant quantization system utilizing the same number of quantization bits. This decrease in quantization bits can lead to substantial reductions in memory size, hardware expenses, and overall system complexity.

\subsection{Processor Allocation}
Efficiently managing the processing power of the DFRC system is critical. Take radars, for instance, which require significant processing capacity to perform tasks such as determining range, angular location, velocity estimation, and tracking. In this process, the radar sends out a series of pulses, each separated by a set interval known as PRI (Pulse Repetition Interval). After each pulse is transmitted, the radar begins to sample and collect data from $N_c$ pulses, forming what is called a frame. This frame corresponds to a Coherent Processing Interval (CPI), which serves as the timeframe for radar processing to commence.

In contrast, communication tasks, especially those involving live video or audio, necessitate more frequent access to the processing unit. Consequently, a dynamic allocation strategy becomes crucial, giving priority to communication processing while periodically releasing resources for radar processing during predefined intervals. This approach ensures that both radar and communication tasks are handled efficiently within the DFRC system.

\subsection{Bandwidth Allocation}
Resource allocation in DFRC systems poses a critical challenge due to their reliance on fixed-bandwidth waveforms for transmission. The bandwidth allocation in DFRC systems influences two critical aspects: range resolution (for sensing) and data throughput (for communication). Achieving the optimal equilibrium in bandwidth allocation becomes imperative to fulfill the unique demands of sensing and communication within DFRC systems, all while upholding the stringent QoS standards in place.

There is significant research interest in studying the radar sensing properties of multi-carrier waveforms, known for their ability to achieve high communication rates \cite{10036975}. Frequency-hopping multi-carrier waveforms were used in \cite{9241739,8645212}. In this approach, the available bandwidth is divided into multiple sub-bands, and a randomly chosen subset of these sub-bands is paired with a transmit antenna for each channel use. However, this method utilizes only a portion of the available bandwidth, degrading target range resolution and communication rate. Orthogonal Frequency-Division Multiplexing (OFDM) waveforms for DFRC systems are explored in \cite{7970102,9420261,9166743}. In these studies, non-overlapping subcarriers are allocated to transmit antennas in each channel use. Contrary to frequency-hopping methods, OFDM approaches make use of the entire available bandwidth, leading to enhanced range resolution and communication rate. However, due to the restriction of each antenna to transmit on specific subcarriers, the communication bandwidth is not fully exploited.

A bandwidth-efficient DFRC system employing OFDM waveforms is proposed in \cite{10036975}. Unlike the OFDM approaches discussed previously, the available subcarriers in this approach are categorized into two groups: shared and private. On shared subcarriers, all antennas can transmit simultaneously, whereas on private subcarriers, only one antenna can transmit at a time. Subcarrier sharing enables a high communication rate. To address the coupling of transmitted symbols and sensing parameters in the target echoes, arising from the shared use of subcarriers, a low complexity target estimation approach is proposed in \cite{10036975}.

\subsection{Beampattern Control}
The beampattern of transmitted signals holds great significance in DFRC systems. For sensing purposes, the strength of the reflected signal from the target relies on the directionality of the transmitted signal toward the target. Therefore, transmitted signals directed at the target must possess sufficient power. In contrast, signals directed towards end-users primarily serve the purpose of information transmission, allowing for lower power signals in their direction, thereby prioritizing energy efficiency while upholding QoS standards.

\section{CHALLENGES AND FUTURE RESEARCH DIRECTIONS}
One of the key challenges for the DFRC system is to address the conflicting beamforming requirements for communications and radar \cite{book2022joint}. For radars, we want to have a wide field-of-view scanning. On the contrary, directive beams are required for communication at mmWave and sub-Terahertz (THz) band. Moreover, efficient BF tracking approaches need to be developed for C\&R operations with mobile nodes. Sensing-assisted communication can play a vital role to address this issue where the sensing capability can be used for tracking the mobile user and forming predictive communication beams. With the availability of large bandwidths (BW) at mmWave and sub-THz bands for DFRC systems, the design of optimal signaling schemes for best utilization of available BW also becomes an important research area \cite{book2022joint}.

Efficient channel estimation for highly directional beams in sparse channels is challenging due to the vast angular search space and limited available measurements. Existing solutions like compressed sensing are computationally complex and sensitive to noise, errors in array response, and off-grid issues \cite{10077114}. In this regard, the sensing capability of DFRC systems can help by identifying the locations of key scatterers in the sparse channel, thus substantially reducing the search space and enhancing estimation accuracy, especially with limited measurements.

Another interesting research direction is the multichannel system in which multiple frequency channels are used during a symbol transmission period. By combining signals from different channels, such systems can offer an overall large BW required for accurate sensing without increasing the communication BW \cite{book2022joint}. The key challenge here is how to remove the distortions from signals across multiple channels before stitching them together for sensing and how to design signals that are easy to concatenate. Readers are referred to \cite{wei2020passive} for an example of a work that deals with multichannel signals for passive radar sensing.

Receiver design for DFRC systems is an emerging research area where little work has been done until now. A novel joint receiver design is proposed in \cite{liu2020joint}. The challenge in receiver design is to discriminate between communication signals and target echoes in the presence of noise.  Machine learning and deep learning approaches can aid in the receiver design of DFRC systems where they can be used to distinguish between the two signals based on their independent statistical characteristics \cite{liu2020joint}. The situation becomes even more complex in the case of bistatic or multistatic DFRC configurations where knowledge of the transmit waveform is not available at the receiver.

Networked sensing presents an important research direction for PMNs \cite{10077114}. By leveraging the collaborative use of multiple TMTs, sensing performance can be enhanced. The primary advantage of networked sensing lies in the macro-diversity achieved through various TMTs, which can boost the reliability of target detection, as the likelihood of all TMT perspectives being obstructed is minimal. However, a significant challenge in networked sensing is tracking of moving targets. Unlike traditional radar, each TMT's coverage is limited and a smooth handover of the sensing task is necessary from one TMT to another. This challenge becomes more complex in networked sensing when transitioning between groups of TMTs with possible overlap between them. While handover is a familiar concept in wireless communication, it differs for sensing due to several factors which are: the challenging requirement of synchronization in networked sensing, the handover criteria is dependent on additional factors including angles and the target's moving direction, and the requirement of "predictive" sensing beams for fast-moving targets \cite{10077114}. To address these challenges, machine learning or data-driven approaches may become more crucial. Graph Neural Networks (GNNs) have shown effectiveness in solving various wireless network design problems, and they can also be applied to the handover problem by formulating it as a graph optimization problem \cite{10077114}.

As far as the military radars are concerned, information about their location, system parameters, and capabilities is to be kept confidential so that the adversary is unable to devise countermeasures against these radars. The security of this information may be compromised by sharing the spectrum with communication systems. Some notable works in this direction employing physical layer security are \cite{tong2022secrecy,wen2022joint,10005137,xu2022robust}. Another potential implication of using a joint system for military sensing and communications is that in the event of a military conflict, the destruction of such a system by the enemy can result in the loss of both capabilities (sensing as well as communications) within a particular coverage area, which is of course not a desired scenario.

DFRC system analysis from an information theoretical point of view needs more work as the current information theoretical foundation for joint systems is limited \cite{book2022joint,liu2020joint}. This work is necessary to reveal the performance bounds for the DFRC systems. Information theoretical analysis for DFRC uplink has been presented in \cite{chiriyath2015inner}, however, the downlink channel has to be investigated further.

\section{CONCLUSION}
DFRC is an emerging research area as it can offer a tunable trade-off between C\&R functionalities. It can tackle the spectrum congestion issue and offer numerous advantages in terms of power, size, cost, reduced interference, improved localization accuracy, and beamforming efficiency. Massive MIMO systems operating in mmWave and sub-THz bands are promising candidates for the implementation of DFRC systems as they can fulfill the requirements of accurate sensing and high data rate communications. Hybrid analog-digital beamforming architecture is suitable for DFRC systems as it can offer the advantages of both analog and digital BF while minimizing their drawbacks. The design of an optimal joint waveform that can perform both functionalities of communication and radar sensing, is the fundamental research problem in DFRC system design. Challenges arising from the inherent differences between C\&R need to be addressed to maximally benefit from this joint design approach. This paper provides a comprehensive review of the state-of-the-art research on DFRC systems covering various design aspects in detail as well as a coverage of potential application scenarios and future research directions. It can serve as a good starting point for readers interested in doing research in this particular area.

\bibliographystyle{IEEEtran}
\bibliography{references1}

\end{document}